\documentclass[12pt,a4paper]{article}
\usepackage{amsmath,setspace}
\usepackage[font={small,it}]{caption}
\usepackage[top=1.2in, bottom=1.2in, left=1in, right=1in]{geometry}
\usepackage[title]{appendix}

\usepackage{natbib}
\usepackage{color,soul}

\DeclareCaptionStyle{italic}[justification=centering]{labelfont={bf},textfont={it},labelsep=colon}
\usepackage{graphicx,psfrag,epsf,textcomp}
\usepackage{enumerate, dsfont}
\usepackage{natbib}
\usepackage{url,xcolor} 
\usepackage{booktabs, subfig, bm, paralist,mathpazo,tikz,longtable,microtype}

\usepackage[pdftex,colorlinks=true]{hyperref}
\definecolor{darkblue}{rgb}{0,0,.6}
\hypersetup{citecolor=darkblue,linkcolor=darkblue,urlcolor=darkblue}
\newcommand{\argmax}{\operatornamewithlimits{argmax}}

\newcommand{\blind}{0}

\graphicspath{{plots/}}

\newsavebox\CBox

\begin{document}

\def\spacingset#1{\renewcommand{\baselinestretch}{#1}\small\normalsize} \spacingset{1}

\if0\blind
{
  \title{\bf Synergy in fertility forecasting: \mbox{Improving forecast accuracy through model averaging}}
    \author{
    Han Lin Shang\footnote{Address: Department of Actuarial Studies and Business Analytics, Level 7, 4 Eastern Road, Macquarie University, NSW 2109, Australia; Email: hanlin.shang@mq.edu.au; ORCID: \url{https://orcid.org/0000-0003-1769-6430}}\\
    Department of Actuarial Studies and Business Analytics \\
 Macquarie University \\
\\
 Heather Booth\footnote{Address: School of Demography, Research School of Social Sciences, Australian National University, Canberra, ACT 2601, Australia; Telephone: +61(2) 6125 4062; Email: heather.booth@anu.edu.au; ORCID: \url{https://orcid.org/0000-0002-8356-0534}} \\
 School of Demography\\
Australian National University\\
 }
  \maketitle
} \fi

\if1\blind
{
  \title{\bf Synergy in fertility forecasting: \mbox{Improving forecast accuracy through model averaging}}
  \maketitle
} \fi

\bigskip
\begin{abstract}
Accuracy in fertility forecasting has proved challenging and warrants renewed attention. One way to improve accuracy is to combine the strengths of a set of existing models through model averaging. The model-averaged forecast is derived using empirical model weights that optimise forecast accuracy at each forecast horizon based on historical data. We apply model averaging to fertility forecasting for the first time, using data for 17 countries and six models. Four model-averaging methods are compared: frequentist, Bayesian, model confidence set, and equal weights. We compute individual-model and model-averaged point and interval forecasts at horizons of one to 20 years. We demonstrate gains in average accuracy of 4-23\% for point forecasts and 3-24\% for interval forecasts, with greater gains from the frequentist and equal-weights approaches at longer horizons. Data for England \& Wales are used to illustrate model averaging in forecasting age-specific fertility to 2036. The advantages and further potential of model averaging for fertility forecasting are discussed. As the accuracy of model-averaged forecasts depends on the accuracy of the individual models, there is ongoing need to develop better models of fertility for use in forecasting and model averaging. We conclude that model averaging holds considerable promise for the improvement of fertility forecasting in a systematic way using existing models and warrants further investigation. 

\vspace{.2in}
\end{abstract}

\newpage
\spacingset{2}

\section{Introduction}

Fertility forecasts are a vital element of population and labour force forecasts, and accurate fertility forecasting is essential for government policy, planning and decision-making regarding the allocation of resources to multiple sectors, including maternal and child health, childcare, education and housing. Though the level of fertility in industrialised countries has at times been of major concern \citep[e.g.,][]{Kuczynski37, Booth86,Longman04}, models for forecasting fertility have been developed relatively recently \citep[for an earlier review, see][]{Booth06}. Indeed, many of the models used in fertility forecasting were originally designed to smooth age-specific fertility rates or complete the incomplete experience of a single cohort, and relatively few have been used in longer-term forecasting of period fertility rates; further, many are deterministic providing no indication of uncertainty \citep{BLM18}. 

Stochastic models for forecasting period age-specific fertility rates (ASFRs) make use of various time-series extrapolation models and generally involve extensive simulation to take account of covariance in the estimation of uncertainty. The most straightforward approach uses univariate time-series models \citep[see, e.g.,][]{BJR08} to extrapolate the trend at each age. The main limitation of this approach is inconsistency among age-specific forecasts, possibly producing an implausible age pattern of future fertility. To remedy this,  the age pattern is first modelled, and its parameters are then forecast. Parametric models used in forecasting include the beta, gamma, double exponential and Hadwiger functions \citep[see, e.g.,][]{TBL+89, Congdon90, Congdon93, KMR93, KP00}, while semi-parametric models include the Coale-Trussell and Relational Gompertz models \citep[see, e.g.,][]{CT74, Brass81, Murphy82, Booth84, ZWM+00}. The use of these models is variously limited by parameter un-interpretability, over-parameterization and the need for vector autoregression. Structural change also limits their utility, especially where vector autoregression is involved \citep{Booth06}.

Nonparametric methods use a dimension-reduction technique, such as principal components analysis, to linearly transform ASFRs to extract a series of time-varying indices to be forecast \citep[see, e.g.,][]{BB87, Bell92, Lee93, HU07, MGC13}. This approach parallels the Lee-Carter model and its many variants and extensions in mortality forecasting \citep[see, e.g.,][]{SBH11}, but has received far less attention. Other contributions include state-space, Bayesian and stochastic diffusion approaches \citep[see, e.g.,][]{RR10,MG13,SGM+14}.
 
The strengths and weaknesses of fertility forecasting models have not been thoroughly evaluated. Noting the absence of guidance on model choice,  \cite{BLM18} compared the point and interval forecast accuracy of 20 major models for fertility forecasting with 162 variants. In the context of completing cohort fertility, their evaluation found that only four methods were consistently more accurate than the constant (no change) model and, among these, complex Bayesian models did not outperform simple extrapolative models. These findings were mostly universal. In earlier research, \citet{Shang12a} compared several models for fertility forecasting for point and interval forecast accuracies, finding the weighted Hyndman-Ullah model (see Section~\ref{sec:models}) to be marginally more accurate on both counts. While this research is useful in identifying models that perform well based on extensive data, such models may not perform well in every circumstance. Moreover, there is undoubtedly scope for fertility forecasting improvement.

In other areas of forecasting, model averaging \citep[][]{BG69, Dickinson75, Clemen89} has been employed to improve point and interval forecast accuracy. However, with the exceptions of \cite{Shang12} and \cite{SH18}, model averaging has been neglected in forecasting demographic rates. This paper aims to empirically assess the extent to which forecast accuracy can be improved through model averaging in the context of age-specific  fertility forecasting. Our assessment covers 17 countries with varied fertility experience and is based on data series extending back to 1950 or before. We consider six selected models and four methods for selecting model averaging weights.  

The structure of this article is as follows. In Section~\ref{sec:3}, we introduce model averaging and briefly present the four methods for selecting weights; the technical details of these methods are given in Appendix~\ref{sec:MA_tech}. The data, six selected models and study design are described in Section~\ref{sec:4}, and technical details of the six models appear in Appendix~\ref{sec:appendix_B}. Based on the accuracy measures discussed in Appendix~\ref{sec:5}, in Section~\ref{sec:6}, we evaluate the point and interval forecast accuracies of the six models and of the four model-averaged forecasts. In Section~\ref{sec:8}, we provide an example of model averaging in forecasting age-specific fertility, using data for England \& Wales. Finally, the discussion appears in Section~\ref{sec:9}. 

\section{Model averaging}\label{sec:3}

The idea of model averaging has been often studied in statistics, dating back to the seminal work by \cite{BG69}. A flurry of articles then appeared dedicated to the topic; see \cite{Clemen89} for a review from a frequentist viewpoint and \cite{HMR+99} for a review from a Bayesian perspective. More recent developments in model averaging are collected in the monograph by \cite{CH08}. In demographic forecasting, there has been limited usage, but notable exceptions include \cite{SS95}, \cite{Ahlburg98, Ahlburg01} and \cite{Sanderson98} in the context of census tract forecasting.

In essence, the model-averaging approach combines forecasts from a set of two or more models. Because these models may reflect different assumptions, model structures and degrees of model complexity, it is expected that better forecast accuracy can be achieved through averaging. The forecasts from each model are averaged using weights that are specific to each forecast horizon. This is designed to achieve accuracy in the year of the forecast horizon, and not in all years up to and including the horizon year. These model weights are applied at all ages. 

In all model-averaging methods, the model-averaged point forecast for horizon $h$ is computed as the weighted mean: 
\begin{equation*}
\widehat{y}_{n+h|n,\text{M}}=\sum_{\ell=1}^{L}w_{\ell}\widehat{y}_{n+h|n,M_{\ell}},
\end{equation*}
where $M$ represents the model average, $M_{\ell}$, for $\ell=1,\dots,L$, represents the individual models, $\widehat{y}_{n+h|n,M_{\ell}}$ represents the point forecast at horizon $h$ obtained from model $\ell$; $\widehat{y}_{n+h|n,\text{M}}$ represents the model averaged forecast; and $(w_1, w_2, \dots, w_{L})$ are empirical point forecast weights that sum to 1.

For the model-averaged point forecast $\widehat{y}_{n+h|n,\text{M}}$,  its prediction interval is constructed from its variance assuming a normal distribution. Following \cite{BA02,BA04}, the unconditional variance of the estimator $\tilde{y}_{n+h|n, \text{M}}$ based on the models weighted by empirical interval forecast weights is given by
\begin{equation*}
  \widehat{\text{Var}}(\tilde{y}_{n+h|n, \text{M}}) = \left\{\sum^{L}_{i=1}\tilde{w}_{\ell}\left[\widehat{\text{Var}}(\widehat{y}_{n+h|n,M_{\ell}})+(\widehat{y}_{n+h|n,M_{\ell}}-\widehat{y}_{n+h|n, \text{M}})^2\right]^{\frac{1}{2}}\right\}^2,
\end{equation*}
where $\tilde{w}_{\ell}$ represents the interval forecast weight under model $M_{\ell}$, $\tilde{y}_{n+h|n, \text{M}}$ represents the model-averaged point forecast obtained using interval forecast weights. The first term inside the square brackets measures the variance conditional on model $M_{\ell}$, while the second term measures the squared bias. 

The crucial ingredient in model averaging is the empirical weights. The computation of the point forecast weights and interval forecast weights is based on measures of forecast accuracy. The derivation of weights differs by model-averaging method. We employ four methods chosen on the basis of their ability to perform well in mortality forecasting \citep{Shang12, SH18}. These four methods for computing the weights are briefly described in the following sections. Further details are given in Appendix~\ref{sec:MA_tech}.

It should be noted that in this study, the measures of accuracy are further averaged as part of the study design (see Section~\ref{sec:4.2}), increasing the reliability and stability of the weights.

\subsection{A frequentist approach}\label{sec:3.1}

Under the frequentist approach employed, the accuracy of a point forecast is measured by mean absolute forecast error (MAFE) while the accuracy of an interval forecast is measured by the mean interval score. These measures are described in Appendix~\ref{sec:5}.

In the simple case of a single point forecast for a particular horizon $h$, MAFE is averaged over age for the year $n+h$, and the weight is taken to be equal to the inverse of MAFE. Weights are obtained for the set of forecasting models and standardised to sum to 1, giving the standardised weights $(w_1, w_2, \dots, w_L)$. Similarly, interval forecast accuracy is based on the mean interval score for year $n+h$, and the weight is equal to its inverse. For the set of models, weights are standardised to achieve proportionality, giving the standardised weights $(w_1, w_2, \dots, w_L)$. Further averaging of MAFE and the mean interval score takes place as a result of study design, before weight calculation.

\subsection{A Bayesian approach}\label{sec:3.2}

In Bayesian model averaging, a single set of model weights is used for the point and interval forecasts. The weights are derived from the Bayesian information criterion (BIC) and are thus proportional to model goodness-of-fit and incorporate a penalty for more parameters. Further details appear in Appendix~\ref{sec:MA_tech}.

\subsection{Model confidence set (MCS)}

To examine statistical significance among the set of models and select a set of superior models, we consider the MCS procedure. The MCS procedure proposed by \cite{HLN11} consists of a sequence of tests of the hypothesis of equal predictive ability (EPA) of two models, eliminating the worse-performing model at each step and resulting in a smaller set of superior models for which the hypothesis is universally accepted. After determining this set of superior models, their forecasts are averaged without weights.

\subsection{Equal weights}\label{sec:equal_weight}

As a baseline approach, we also consider assigning equal weights to all six models. Prior research has shown that including all models in an equal-weights (or unweighted) model can yield substantial gains in forecast accuracy. Several studies provide analytical solutions for the conditions under which equal weights provide more accurate out-of-sample forecasts than regression weights \citep[e.g.,][]{EH75, Davis-Stober11}. Equal weights can be advantageous when models fit the data poorly \citep{Graefe15}.

\section{Data and design}\label{sec:4}

\subsection{Data sets}

The study is based on fertility data for 17 countries. For all but one country, the data were taken from the \cite{Database12}. Australian fertility rates and populations were obtained from Australian Bureau of Statistics (Cat. No. 3105.0.65.001, Table 38); this data set is also available in the \textit{rainbow} package \citep{SH12} in the statistical software R \citep{Team12}.  The data consist of annual fertility rates by single-year age of women aged 15 to 49 and corresponding population  (births to women aged 50 and older are included in births at age 49). The 17 selected countries all have reliable data series commencing in 1950 or earlier, as shown in Table~\ref{tab:1}, and extending to 2011.

\begin{singlespace}
\begin{table}[!htbp]
  \caption{Countries and period of data availability}\label{tab:1}
\tabcolsep 0.08in
  \centering
  \begin{tabular}{@{}llllllllll@{}}\toprule
    Country & Year & & & Country & Year & & & Country & Year \\\toprule
    Australia & 1921-2015 & & & Austria & 1951-2017 & & & Canada & 1921-2011 \\
    Czech Republic & 1950-2016 & & & Denmark & 1916-2016 & & & England \& Wales & 1938-2016 \\
    Finland & 1939-2015 & & & France & 1946-2016 & & & Hungary & 1950-2017 \\
    Japan & 1947-2016 & & & Netherlands & 1950-2016 & & & Scotland & 1945-2016 \\
    Slovakia & 1950-2014 & & & Spain & 1922-2016 & & & Sweden & 1891-2016 \\
    Switzerland & 1932-2016 & & & USA & 1933-2016 \\\toprule
  \end{tabular}
\end{table}
\end{singlespace}

\begin{figure}[!htbp]
\centering
\includegraphics[width=12cm]{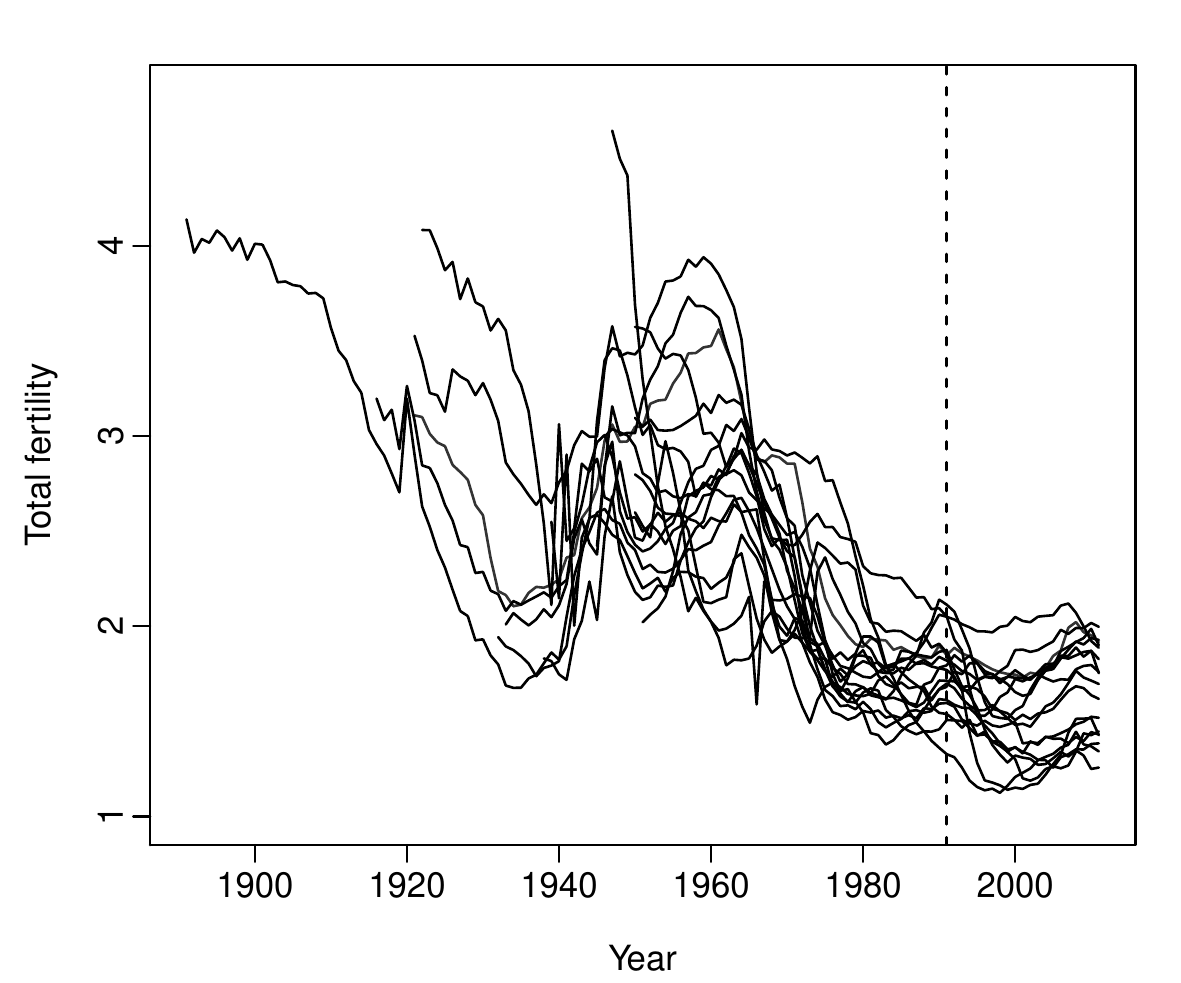}
\caption{Total fertility of 17 selected countries, where the vertical dotted line divides the in-sample period and the holdout sample period} \label{fig:1}
\end{figure}

The overall level of fertility has declined in all countries considered. Figure~\ref{fig:1} shows that the general decline in total fertility (births per woman calculated as the sum of ASFRs in each year) has not been monotonic. It also indicates that the substantial decline in fertility since 1960 has been stalled or reversed since about 2000. While trends in ASFRs may differ from this overall pattern, total fertility provides the average trend.

\subsection{Six selected models}\label{sec:models}

We select six models used in forecasting fertility. These include three functional time-series models for modelling the schedule of ASFRs, where age is treated as a continuum; and three univariate time-series models to model fertility at each age individually. All models are applied to transformed ASFRs so as ensure non-negative forecast rates.

\subsubsection{Functional time-series models}

The three models included here stem from the work of \cite{HU07}. These models use principal component decomposition with time series forecasting of the time parameter, and are similar to the well-known method by \cite{LC92} for modelling and forecasting mortality rates. 

The Hyndman-Ullah (HU) model uses smoothed transformed ASFRs. The functional principal component decomposition produces smooth age parameters and corresponding time parameters. Several components are required to adequately describe the data. The time parameters are each modelled using time series methods, providing the forecast.

Two variants of the HU model are also selected. The robust HU model (HUrob) is robust to outliers. The weighted HU model (HUw) gives greater weight to more recent data in order to reduce the potential gap between the last year of observed data and the first year of the forecast. The HU model and its two variants are described in detail in Appendix~\ref{sec:appendix_B}.

\subsubsection{Univariate time-series models}

The three selected univariate time-series models \citep{BJR08} include two random-walk models and an optimal autoregressive integrated moving average (ARIMA) model, and are applied to transformed ASFRs at each age. They are described in detail in Appendix~\ref{sec:appendix_B}.

The random-walk models include the random walk (RW) and the random walk with drift (RWD). Each model is applied independently to each age-specific rate. While the RW model assumes a constant underlying rate over time with all deviations assumed to be error, the RWD assumes a constant underlying change in the transformed rates at each age.

The optimal ARIMA model is the model best describing the time series of an age-specific fertility rate based on its own past values, including its own lags and lagged forecast errors. The equation is then used to forecast future values of the series.

\subsection{Study design}\label{sec:4.2}

\begin{figure}[!t]
\centering
\includegraphics[width=16.5cm]{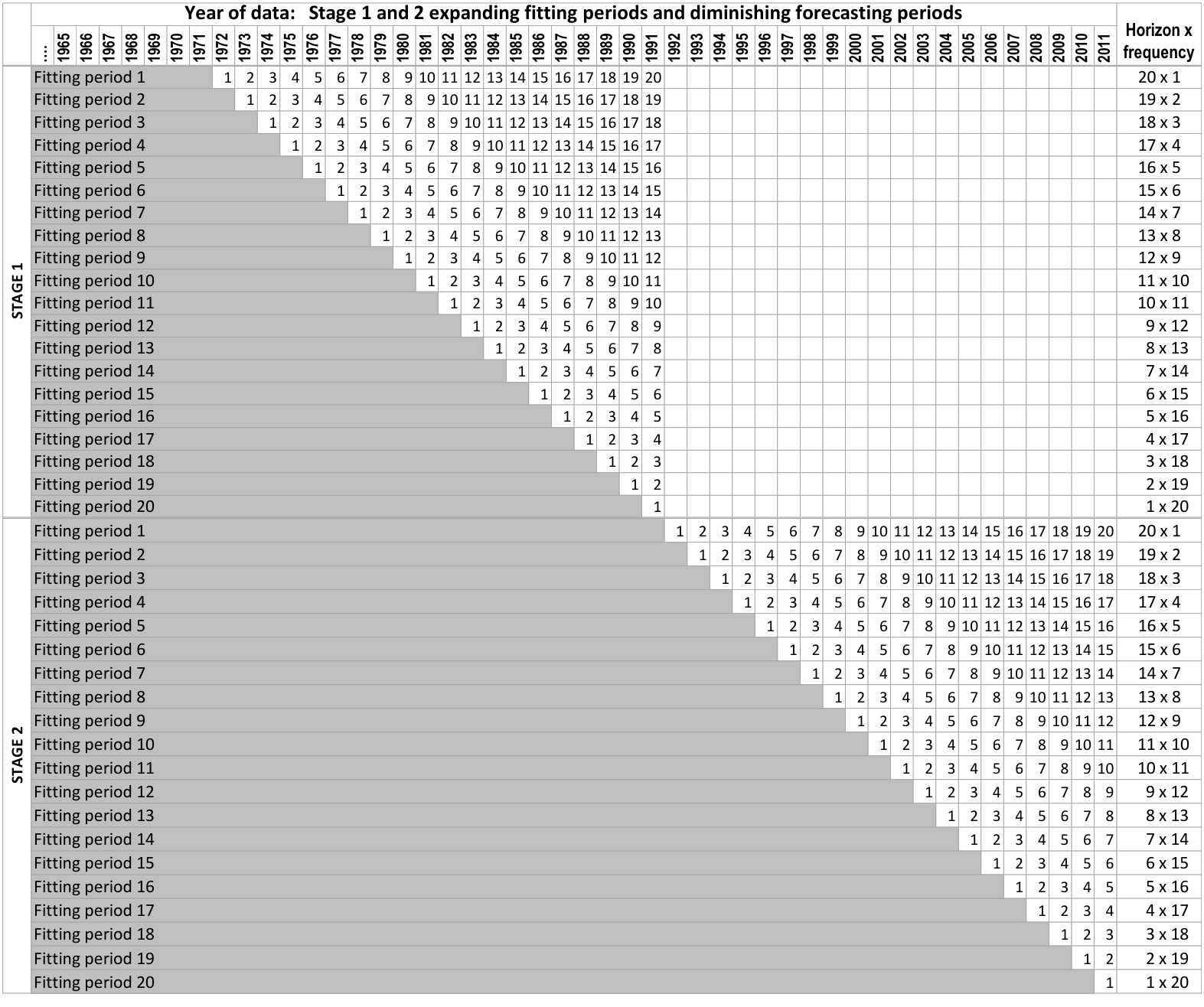}
\caption{Fitting and forecasting periods for weight estimation (stage 1) and model-averaged forecasting (stage 2). The start of the fitting period is determined by data availability or purposive choice.}\label{fig:Fig_2X}
\end{figure}

We implement a two-stage design. At the first stage, the six forecasting models are applied to determine model weights based on forecast errors. At the second stage, the six models and their model averages are evaluated. These two stages use different periods of data as illustrated in Figure~\ref{fig:Fig_2X}. We divide the data for each country into an in-sample period ending in 1991 and a holdout sample period, 1992 to 2011.

\begin{figure}[!htbp]
\centering
\subfloat[One-year-ahead forecast: 1972]
{\includegraphics[width=8.8cm]{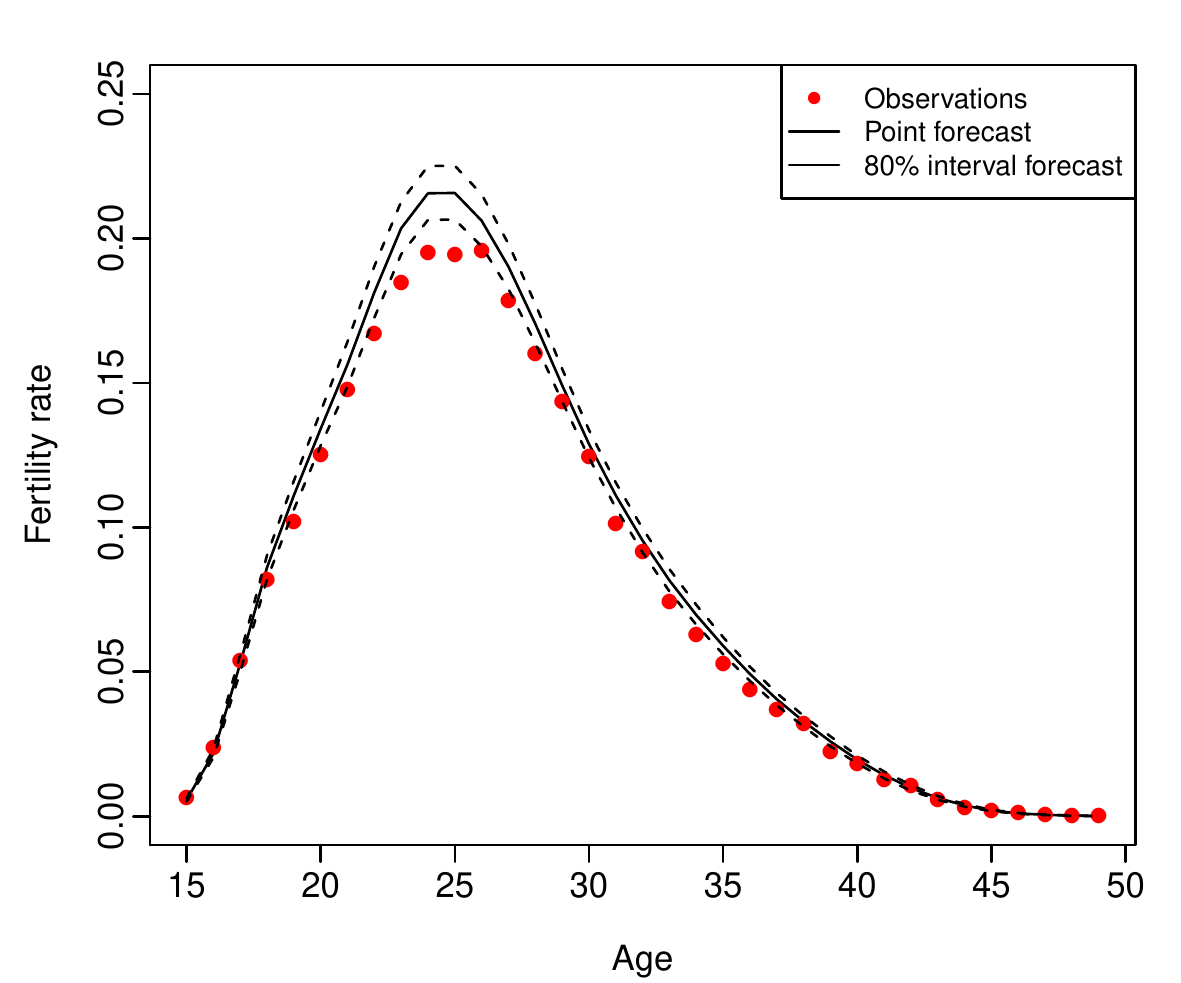}\label{fig:insample1step}}
\qquad
\subfloat[20-year-ahead forecast: 1991]
{\includegraphics[width=8.8cm]{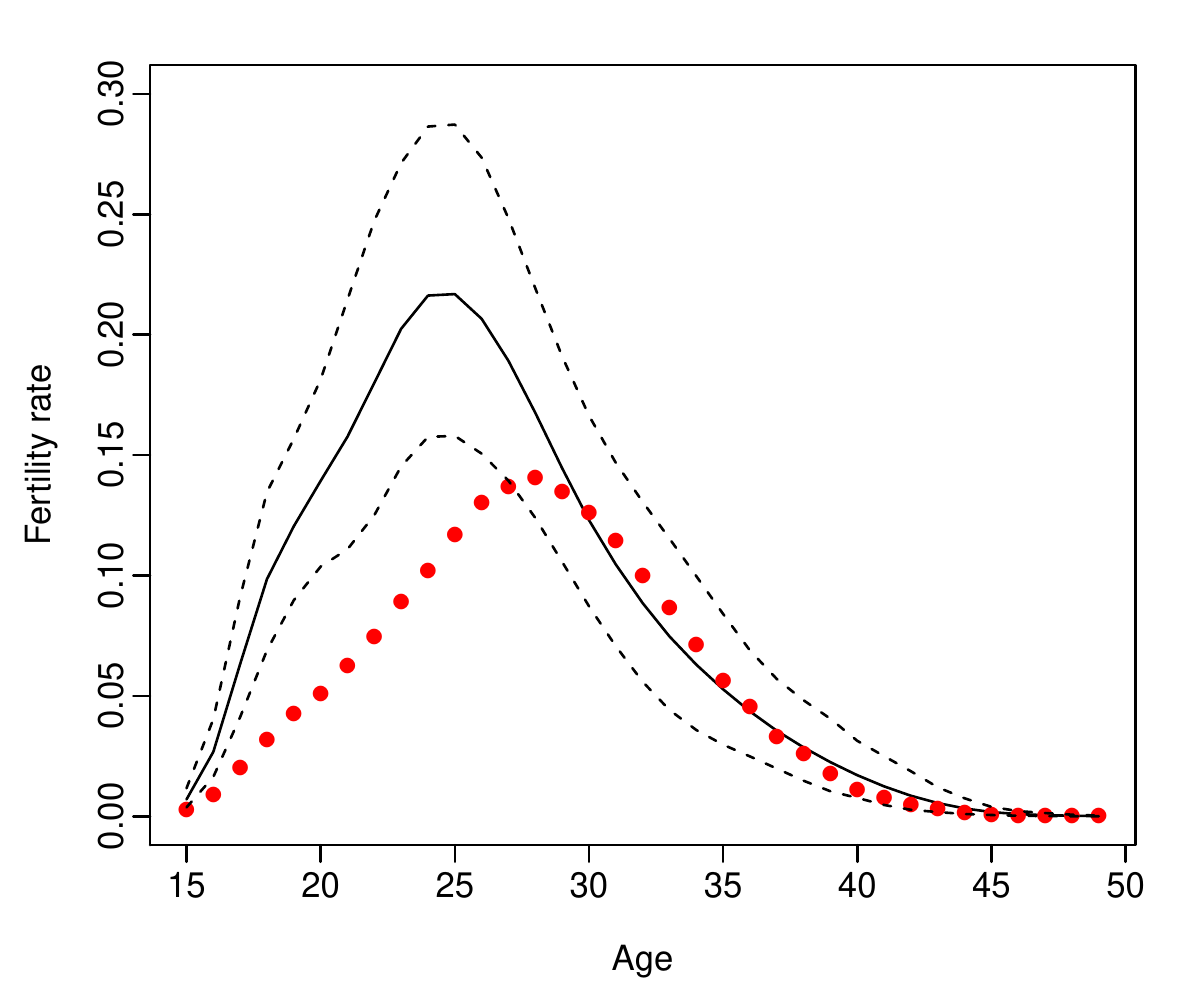}\label{fig:insample20step}}
\caption{Comparison of observed data with one-year-ahead and 20-year-ahead point and interval forecasts of Australian ASFRs based on data for 1921-1971 using the HU model}\label{fig:2}
\end{figure}

The in-sample data are used in the first stage. We further divide these data into a fitting period and a forecasting period. The initial fitting period ended in 1971, and the forecasting period was 1972 to 1991. Using the data in this fitting period, we fit each model, compute one-year-ahead to 20-year-ahead forecasts of 35 ASFRs, and calculate forecast errors or information criterion values by comparing the forecasts with observed data in the relevant year. Figure~\ref{fig:2} provides an example. Then, we expand the fitting period by one year, and compute one-year-ahead to 19-year-ahead forecasts, and again calculate forecast errors or information criterion values. This process is repeated until the fitting period extends to 1990. In so doing, we have 20 sets of one-year-ahead forecast errors, 19 sets of two-year-ahead forecast errors, $\dots$ and one set of 20-year-ahead forecast errors. These forecast errors are used in the calculation of appropriate weights for the model-averaging approach. In practice, the model-averaging weights are derived from forecast errors averaged over horizon-specific forecasts and over the 17 countries in the study. There is thus one set of six weights for each model-averaging method.

These forecast errors are used in the calculation of appropriate weights for the model-averaging approach. In the HUw model where geometrically-decaying weights are used, the decaying parameter, $\lambda$, is estimated from data for 1960 to 1981, for $h=1,\dots,20$. The overlap with the in-sample forecasting period (1970 to 1991) is inevitable given short data series for some countries. 

At the second stage, the entire dataset is used. The initial fitting period ended in 1991, and the forecasting period was 1992 to 2011. An expanding fitting period is again employed, producing forecasts for horizons of 1 to 20 years. Point and interval forecasts are produced for the six forecasting models and for the model-averaged forecasts using each of the four model averaging methods. Forecast errors are again calculated by comparison with observed data and used in the evaluation of point and interval forecast accuracy as measured by MAFE and the mean interval score, respectively.

Implementation of the HU models presented in this paper is straightforward with the readily available R package \textit{demography} \citep{Hyndman11}. Point and interval forecasts based on the RW, RWD and ARIMA models can be obtained via the \verb rwf \ and \verb auto.arima \ functions in the \textit{forecast} package \citep{Hyndman12}. The computational code in R for the entire analysis is available upon request.

\section{Results}\label{sec:6}

\subsection{Point forecast accuracy}\label{sec:61}

In Table~\ref{tab:2}, we present the out-of-sample point forecast accuracy based on one-year-ahead to 20-year-ahead MAFEs, which are averaged over ages, years in the forecasting period and countries. Irregularities at longer horizons are due to two factors: the small number of forecasts involved (see Figure 2) and the fact that they are based entirely on fitting periods ending in 1986 to 1990, a period of changing trends in several countries (see Figure~\ref{fig:1}).

\begin{table}[!htbp]
\centering
\tabcolsep 0.24cm
\caption{MAFEs ($\times 100$) for ASFRs for one-year-ahead to 20-year-ahead point forecasts by model/method and forecast horizon. The MAFEs are averaged over ages, years in the forecasting period and countries}\label{tab:2}
\begin{tabular}{@{}rcccccccccc@{}}
  \toprule
 & \multicolumn{6}{c}{Six models} & \multicolumn{4}{c}{Model average} \\
 Horizon & HU & HUrob & HUw & RW & RWD & ARIMA & Frequentist & Bayesian & MCS & Equal \\ 
  \midrule
  1 & 0.16 & 0.16 & 0.14 & 0.16 & 0.17 & 0.15 & 0.14 & 0.16 & 0.16 & 0.14 \\ 
  2 & 0.25 & 0.25 & 0.22 & 0.27 & 0.29 & 0.24 & 0.23 & 0.27 & 0.26 & 0.23 \\ 
  3 & 0.34 & 0.34 & 0.31 & 0.38 & 0.41 & 0.34 & 0.32 & 0.38 & 0.37 & 0.32 \\ 
  4 & 0.43 & 0.43 & 0.40 & 0.49 & 0.53 & 0.44 & 0.40 & 0.48 & 0.47 & 0.40 \\ 
  5 & 0.51 & 0.52 & 0.48 & 0.59 & 0.64 & 0.54 & 0.49 & 0.58 & 0.56 & 0.49 \\ 
  6 & 0.60 & 0.60 & 0.57 & 0.69 & 0.75 & 0.63 & 0.57 & 0.67 & 0.66 & 0.57 \\ 
  7 & 0.69 & 0.69 & 0.65 & 0.79 & 0.87 & 0.74 & 0.66 & 0.77 & 0.75 & 0.66 \\ 
  8 & 0.78 & 0.78 & 0.76 & 0.88 & 0.98 & 0.85 & 0.75 & 0.86 & 0.84 & 0.75 \\ 
  9 & 0.87 & 0.87 & 0.86 & 0.98 & 1.08 & 0.96 & 0.83 & 0.96 & 0.94 & 0.83 \\ 
  10 & 0.96 & 0.96 & 0.96 & 1.07 & 1.19 & 1.06 & 0.92 & 1.05 & 1.02 & 0.92 \\ 
  11 & 1.05 & 1.05 & 1.05 & 1.16 & 1.29 & 1.16 & 1.00 & 1.14 & 1.11 & 1.00 \\ 
  12 & 1.14 & 1.13 & 1.13 & 1.25 & 1.39 & 1.25 & 1.07 & 1.23 & 1.20 & 1.07 \\ 
  13 & 1.23 & 1.23 & 1.22 & 1.34 & 1.49 & 1.36 & 1.16 & 1.32 & 1.29 & 1.15 \\ 
  14 & 1.33 & 1.32 & 1.30 & 1.44 & 1.59 & 1.46 & 1.25 & 1.42 & 1.38 & 1.24 \\ 
  15 & 1.44 & 1.42 & 1.43 & 1.53 & 1.69 & 1.56 & 1.33 & 1.51 & 1.48 & 1.33 \\ 
  16 & 1.53 & 1.53 & 1.51 & 1.63 & 1.80 & 1.66 & 1.44 & 1.61 & 1.58 & 1.43 \\ 
  17 & 1.63 & 1.63 & 1.58 & 1.74 & 1.91 & 1.76 & 1.54 & 1.71 & 1.70 & 1.54 \\ 
  18 & 1.76 & 1.74 & 1.78 & 1.85 & 2.03 & 1.88 & 1.67 & 1.80 & 1.83 & 1.67 \\ 
  19 & 1.88 & 1.88 & 2.01 & 1.96 & 2.14 & 2.02 & 1.79 & 1.93 & 1.94 & 1.78 \\ 
  20 & 2.00 & 2.00 & 2.08 & 2.07 & 2.24 & 2.19 & 1.88 & 2.05 & 2.02 & 1.88 \\\midrule
  Median & 1.00 & 1.00 & 1.01 & 1.11 & 1.24 & 1.11 & 0.96 & 1.10 & 1.07 & 0.96 \\ 
\bottomrule
\end{tabular}
\end{table}

As shown in Table~\ref{tab:2}, we find that gains in forecast accuracy due to model averaging have been achieved at longer $(h>7)$ rather than shorter forecast horizons. The median value of MAFE is $0.0096$ for the model-averaged forecasts with weights selected by the frequentist approach and equal weights, which indicates gains in accuracy over the six models of 4 to $23\%$. By contrast, the model averaging methods based on the Bayesian approach and MCS do not perform well.  

\subsection{Interval forecast accuracy}\label{sec:62}

In Table~\ref{tab:3}, we present the mean interval scores for the one-year-ahead to 20-year-ahead forecasts. As expected, the mean interval score increases with the forecast horizon, reflecting the loss of interval forecast accuracy as the horizon increases. For short horizons, the model-averaged interval forecasts perform less well than individual models, and the Bayesian and MCS approaches are briefly superior. For horizons of 7 or more years, however, the frequentist and equal weights approaches consistently out-perform the individual models and the more complex model-averaging approaches. Across all horizons, the smallest median $(5.11)$ of the mean interval scores $(\times 100)$ occurs for the model-averaged forecasts based on the frequentist and equal weights approaches. 

\begin{table}[!htbp]
\caption{Mean interval score $(\times 100)$ for ASFRs for one-year-ahead to 20-year-ahead interval forecasts by model/method and forecast horizon. The mean interval scores are averaged over ages, years in the forecasting period and countries}\label{tab:3}
\centering
\tabcolsep 0.22cm
\begin{tabular}{@{}rrrrrrrrrrr@{}}
  \toprule
 & \multicolumn{6}{c}{Six models} & \multicolumn{4}{c}{Model average} \\
 Horizon & HU & HUrob & HUw & RW & RWD & ARIMA & Frequentist & Bayesian & MCS & Equal  \\ 
  \midrule
1 & 0.84 & 0.85 & 0.80 & 0.90 & 0.90 & 0.82 & 1.26 & 0.96 & 1.12 & 1.25 \\ 
  2 & 1.25 & 1.26 & 1.26 & 1.42 & 1.43 & 1.30 & 1.58 & 1.47 & 1.65 & 1.59 \\ 
  3 & 1.76 & 1.77 & 1.87 & 1.93 & 1.99 & 1.85 & 1.98 & 1.92 & 2.09 & 1.99 \\ 
  4 & 2.22 & 2.24 & 2.59 & 2.43 & 2.58 & 2.38 & 2.35 & 2.41 & 2.48 & 2.35 \\ 
  5 & 2.68 & 2.71 & 3.09 & 2.92 & 3.18 & 2.93 & 2.75 & 2.89 & 2.89 & 2.74 \\ 
  6 & 3.14 & 3.16 & 3.58 & 3.40 & 3.79 & 3.49 & 3.15 & 3.38 & 3.28 & 3.14 \\ 
  7 & 3.62 & 3.63 & 3.96 & 3.90 & 4.44 & 4.08 & 3.55 & 3.90 & 3.72 & 3.54 \\ 
  8 & 4.08 & 4.07 & 4.51 & 4.39 & 5.08 & 4.66 & 4.01 & 4.40 & 4.16 & 4.01 \\ 
  9 & 4.57 & 4.54 & 4.95 & 4.89 & 5.74 & 5.26 & 4.43 & 4.92 & 4.64 & 4.42 \\ 
  10 & 5.04 & 5.01 & 5.30 & 5.39 & 6.40 & 5.88 & 4.89 & 5.44 & 5.11 & 4.89 \\ 
  11 & 5.52 & 5.47 & 5.62 & 5.89 & 7.08 & 6.51 & 5.33 & 5.95 & 5.53 & 5.33 \\ 
  12 & 6.05 & 5.99 & 6.30 & 6.39 & 7.78 & 7.16 & 5.80 & 6.49 & 6.02 & 5.81 \\ 
  13 & 6.57 & 6.50 & 6.79 & 6.89 & 8.48 & 7.87 & 6.29 & 7.04 & 6.54 & 6.30 \\ 
  14 & 7.21 & 7.08 & 7.35 & 7.41 & 9.18 & 8.60 & 6.73 & 7.65 & 7.09 & 6.76 \\ 
  15 & 7.81 & 7.68 & 8.16 & 7.95 & 9.91 & 9.39 & 7.38 & 8.29 & 7.65 & 7.42 \\ 
  16 & 8.34 & 8.23 & 8.50 & 8.49 & 10.64 & 10.18 & 7.88 & 8.97 & 8.28 & 7.93 \\ 
  17 & 8.94 & 8.78 & 9.10 & 9.06 & 11.40 & 11.04 & 8.47 & 9.55 & 8.85 & 8.57 \\ 
  18 & 9.82 & 9.67 & 9.34 & 9.63 & 12.17 & 12.02 & 9.26 & 10.16 & 9.67 & 9.38 \\ 
  19 & 10.51 & 10.24 & 10.60 & 10.18 & 12.89 & 13.22 & 9.83 & 10.74 & 10.20 & 9.94 \\ 
  20 & 11.20 & 10.86 & 11.66 & 10.68 & 13.60 & 14.45 & 10.81 & 11.31 & 10.72 & 10.97 \\ \midrule
  Median & 5.28 & 5.24 & 5.46 & 5.64 & 6.74 & 6.19 & 5.11 & 5.70 & 5.32 & 5.11 \\ 
\bottomrule
\end{tabular}
\end{table}

\section{Model averaging in practice}\label{sec:8}

To illustrate the application of model averaging to potentially achieve greater forecast accuracy, we use data for England \& Wales for 1938 to 2016 and produce frequentist model-averaged point and interval forecasts for horizons of one year and 20 years, based on the six models. At the first stage, horizon-specific weight estimation is based on fitting periods commencing in 1938 and ending successively in 1996 to 2015, with the corresponding forecasting periods starting in 1997 to 2016. In Table~\ref{tab:4}, we present empirical weights for producing point and interval frequentist model-averaged forecasts. These weights vary little across models, indicating that the six models do not differ appreciably in accuracy. Further, for each model the point forecast weights and interval forecast weights are roughly equal, showing that the point and interval forecasts are compatible with each other. The fact that the weights differ little across horizons indicates that, in this example, the models do not gain or lose relative advantage over the forecasting period. 

\begin{table}[!htbp]
\centering
\tabcolsep 0.065in
\caption{Point-forecast and interval-forecast frequentist model-averaging weights for fertility rates in England \& Wales by model and horizon}\label{tab:4}
\begin{tabular}{@{}rrrrrrrrrrrrrr@{}}
\toprule
	& \multicolumn{6}{c}{Point-forecast weights} & & \multicolumn{6}{c}{Interval-forecast weights} \\
 Horizon & HU & HUrob & HUw & RW & RWD & ARIMA & & HU & HUrob & HUw & RW & RWD & ARIMA \\ 
\midrule
1 & 0.16 & 0.16 & 0.19 & 0.16 & 0.16 & 0.17 & & 0.17 & 0.17 & 0.18 & 0.16 & 0.16 & 0.16 \\  
  2 & 0.17 & 0.17 & 0.19 & 0.15 & 0.15 & 0.17 & & 0.18 & 0.18 & 0.18 & 0.15 & 0.15 & 0.16 \\  
  3 & 0.17 & 0.17 & 0.19 & 0.15 & 0.15 & 0.17 & & 0.18 & 0.18 & 0.18 & 0.15 & 0.15 & 0.16 \\ 
  4 & 0.18 & 0.18 & 0.18 & 0.15 & 0.15 & 0.16 & & 0.18 & 0.18 & 0.18 & 0.15 & 0.15 & 0.16 \\ 
  5 & 0.18 & 0.18 & 0.17 & 0.15 & 0.15 & 0.16 & & 0.18 & 0.18 & 0.18 & 0.16 & 0.15 & 0.16 \\  
  6 & 0.18 & 0.18 & 0.17 & 0.16 & 0.15 & 0.16 & & 0.18 & 0.18 & 0.18 & 0.16 & 0.15 & 0.16 \\ 
  7 & 0.18 & 0.18 & 0.18 & 0.16 & 0.15 & 0.15 & & 0.18 & 0.18 & 0.18 & 0.16 & 0.14 & 0.16 \\ 
  8 & 0.19 & 0.18 & 0.17 & 0.16 & 0.15 & 0.15 & & 0.18 & 0.18 & 0.18 & 0.16 & 0.14 & 0.16 \\ 
  9 & 0.19 & 0.19 & 0.17 & 0.16 & 0.15 & 0.15 & & 0.18 & 0.18 & 0.17 & 0.16 & 0.14 & 0.16 \\  
  10 & 0.19 & 0.19 & 0.17 & 0.16 & 0.15 & 0.15 & & 0.18 & 0.18 & 0.17 & 0.16 & 0.14 & 0.16 \\  
  11 & 0.19 & 0.19 & 0.16 & 0.16 & 0.15 & 0.15 & & 0.18 & 0.18 & 0.18 & 0.16 & 0.14 & 0.17 \\ 
  12 & 0.19 & 0.19 & 0.16 & 0.16 & 0.15 & 0.15 & & 0.18 & 0.18 & 0.18 & 0.15 & 0.14 & 0.17 \\ 
  13 & 0.19 & 0.19 & 0.16 & 0.17 & 0.15 & 0.15 & & 0.18 & 0.18 & 0.19 & 0.15 & 0.13 & 0.17 \\  
  14 & 0.19 & 0.19 & 0.15 & 0.17 & 0.15 & 0.15 & & 0.18 & 0.18 & 0.19 & 0.15 & 0.13 & 0.17 \\  
  15 & 0.19 & 0.19 & 0.16 & 0.17 & 0.15 & 0.15 & & 0.18 & 0.18 & 0.19 & 0.15 & 0.13 & 0.17 \\  
  16 & 0.18 & 0.18 & 0.19 & 0.16 & 0.15 & 0.15 & & 0.17 & 0.17 & 0.20 & 0.15 & 0.13 & 0.17 \\  
  17 & 0.17 & 0.17 & 0.21 & 0.15 & 0.15 & 0.15 & & 0.17 & 0.17 & 0.19 & 0.15 & 0.13 & 0.18 \\  
  18 & 0.17 & 0.17 & 0.21 & 0.15 & 0.15 & 0.15 & & 0.17 & 0.17 & 0.19 & 0.15 & 0.13 & 0.18 \\  
  19 & 0.16 & 0.16 & 0.23 & 0.15 & 0.15 & 0.15 & & 0.17 & 0.17 & 0.19 & 0.16 & 0.13 & 0.18 \\  
  20 & 0.17 & 0.17 & 0.17 & 0.17 & 0.16 & 0.17 & & 0.17 & 0.17 & 0.19 & 0.16 & 0.13 & 0.18 \\ 
   \bottomrule
\end{tabular}
\end{table}

\begin{figure}[!htbp]
\centering
\subfloat[One-year-ahead forecast: 2017]
{\includegraphics[width=8.6cm]{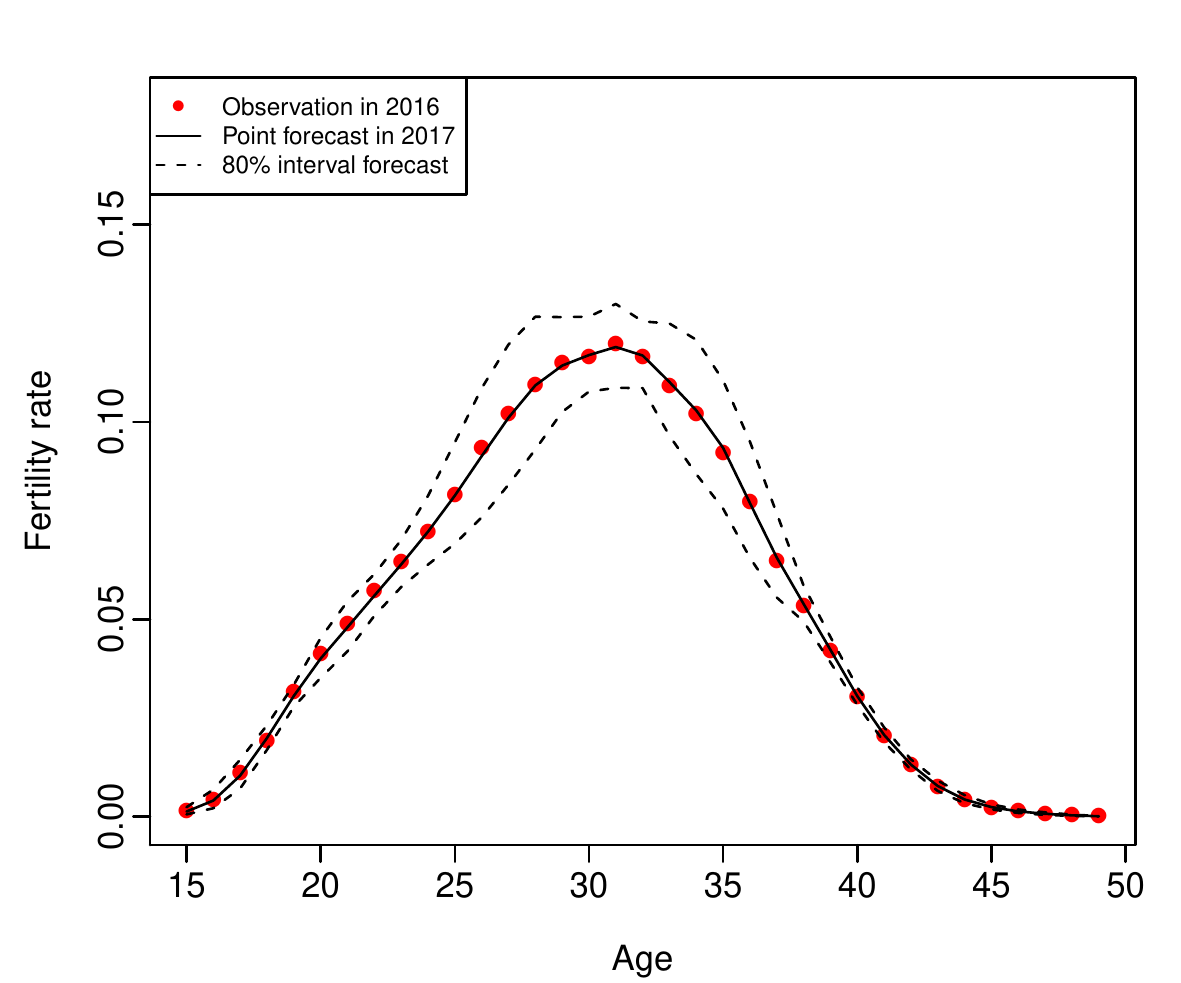}}
\qquad
\subfloat[20-year-ahead forecast: 2036]
{\includegraphics[width=8.6cm]{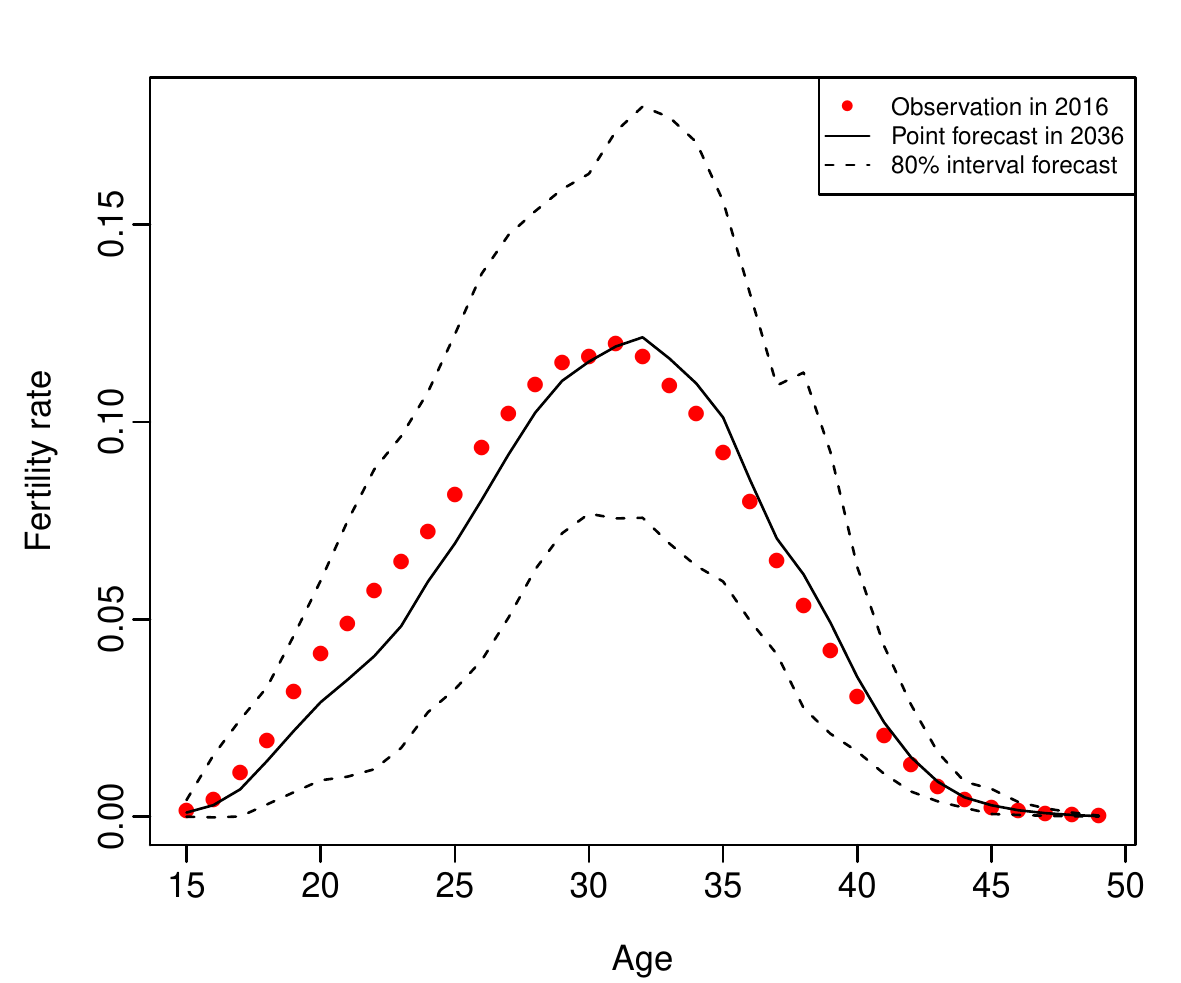}}
\\
\subfloat[One-year-ahead forecast: 2017]
{\includegraphics[width=8.6cm]{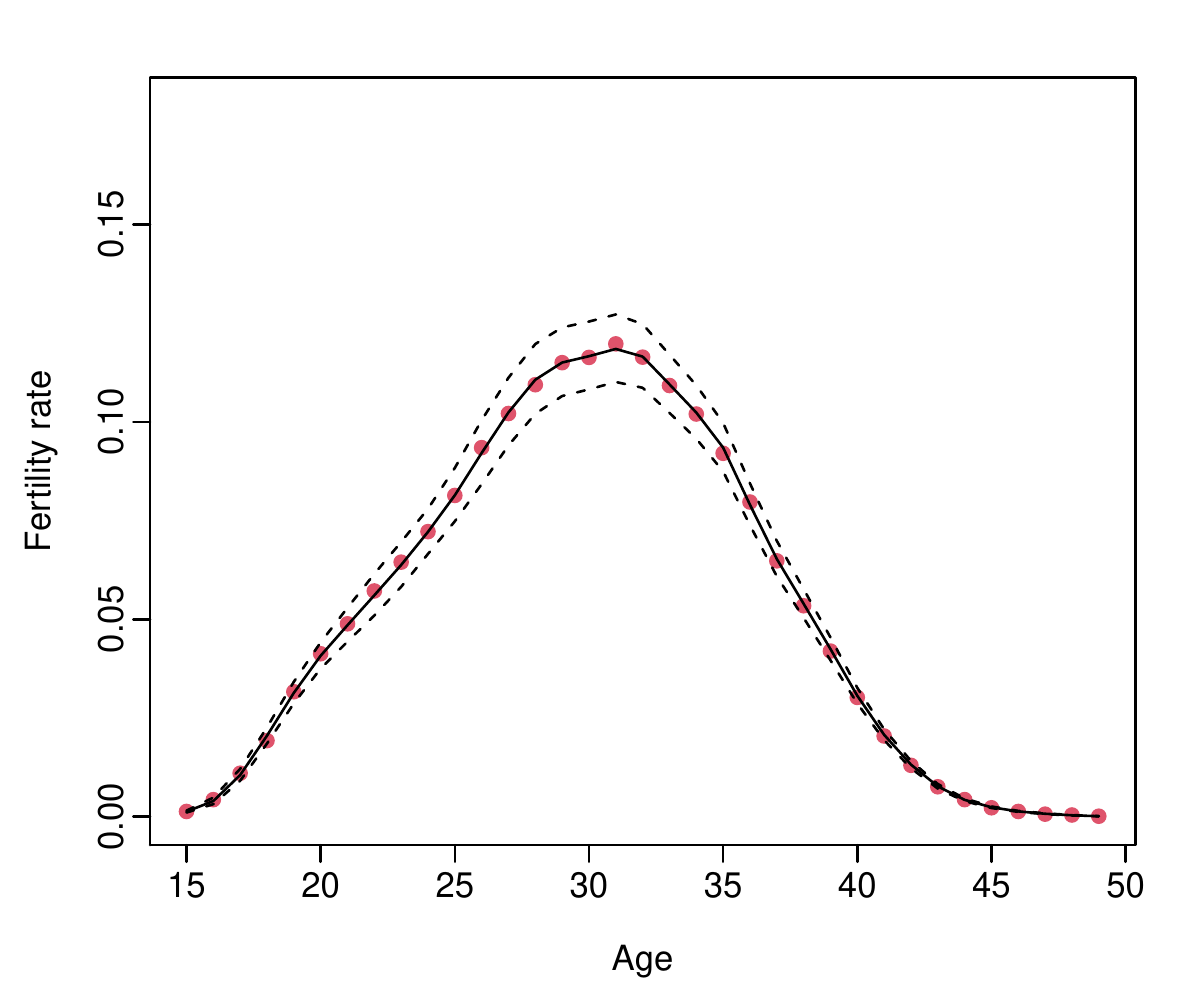}}
\qquad
\subfloat[20-year-ahead forecast: 2036]
{\includegraphics[width=8.6cm]{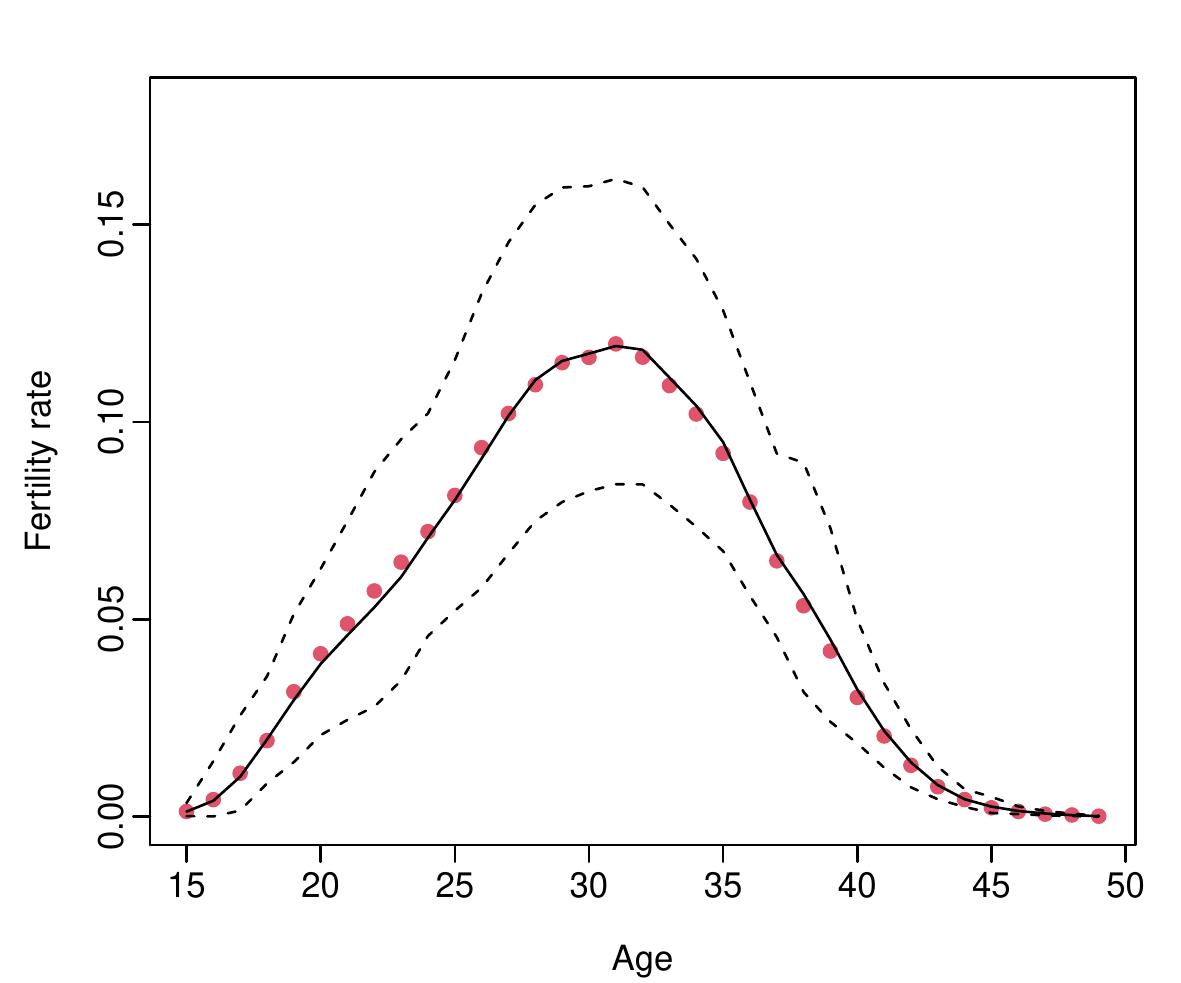}}
\caption{Point and 80\% interval forecasts of age-specific fertility rates in England \& Wales, based on data for 1938 to 2016 using model averaging with frequentist weights (in the first row) and BIC weights (in the second row)}\label{fig:3}
\end{figure}

At the second stage, the fitting period is 1938 to 2016, and the forecasts are produced using the empirical weights obtained at the first stage. The one-year-ahead and 20-year-ahead model-averaged point and interval forecasts of age-specific fertility are shown in Figure~\ref{fig:3}. The 20-year-ahead point forecast portrays a shift in fertility rates to older ages, continuing the previous trend. However, the shift is more pronounced before than after the mode, which changes little, indicating that the tempo effect is nearing the end of its course. Additionally, this forecast substantially reduces the bulge in fertility rates at young ages seen in recent decades \citep{CCH99}.

\section{Discussion}\label{sec:9}

This paper has used computational methods to demonstrate that the accuracy of point and interval forecasts for age-specific fertility can be improved through model averaging. The investigation involved four methods for the empirical determination of the weights to be used in model averaging for both the point forecast and interval forecast. Six models produced the initial set of age-specific fertility forecasts. Among the four model averaging approaches compared, the frequentist and equal-weights approaches performed best on average, over 17 countries and forecast horizons of 1 to 20 years. This evaluation used weights derived from the accuracy of forecasts for 1992 to 2011 based on a long series of observations up to 1991. The frequentist and equal-weights approaches produced gains in point forecast accuracy of 4\% to 23\% over the six individual models. However, the Bayesian and MCS approaches produced reductions of up to 10\%, though some gains were achieved. For the interval forecast, the frequentist and equal-weights approaches also produced the greatest improvement in accuracy, with gains of 3\% to 24\%. Again, the Bayesian and MCS approaches produced the least gains and some losses in forecast accuracy.  

These results hold considerable promise for the improvement of fertility forecasting in a systematic way using existing models. The frequentist approach and the equal weights approach both offer a simple way to implement model averaging. Combined with the findings of \cite{BLM18} that simple forecasting models are as accurate as the most complex, it would seem that improved accuracy can be readily achieved through simple model averaging of simple models. The often cited impediment to the implementation of model averaging, namely complexity, is thus largely removed. To further assist the uptake of model averaging, we look forward to software facilitating its application in fertility forecasting. 

\subsection{Model-averaging approaches}

Among the four approaches for the selection of weights, the frequentist and equal-weights approaches performed much better than the more complex Bayesian and MCS approaches. The superior performance of the frequentist approach can be attributed to the fact that it assigns weights based on in-sample forecast accuracy. Since ASFRs change in limited ways across age and time, it can be expected that a model that produces smaller in-sample point and interval forecast errors will also produce smaller out-of-sample errors. The performance of the equal-weights approach is consistent with previous research \citep{Graefe15}. It is noted that variation in accuracy across models in Tables~\ref{tab:2} and~\ref{tab:3} is somewhat limited, partially explaining the superior performance of the equal-weights approach. The poor performance of the Bayesian approach can be attributed to the fact that the weights are determined based on in-sample goodness of fit rather than in-sample forecast errors, given that model goodness of fit is not a good indicator of forecast accuracy \citep{MHP19}. The poor performance of the MCS approach can be attributed to the 85\% confidence level employed. With a higher confidence level, the MCS method would have selected more accurate models.  

\subsection{Advantages of model averaging}

An essential advantage of model averaging is its synergy. In other words,  improvements in forecast accuracy can be achieved while using existing methods.  Further, in order to achieve optimal accuracy, the 'mix' of the set of models employed is allowed to change over the forecasting period by using horizon-specific weights favouring more accurate methods. Thus, a model that is highly accurate in the short-term but inaccurate in the long term (or vice versa) can be included in the set of models without jeopardising the accuracy of the model-averaged forecast. This allows the forecaster to make use of a wide range of models and approaches including those that might otherwise be regarded as unsuitable because of their horizon-limited accuracy. Moreover, the weights determining the mix of models is empirical, removing an element of subjectivity in model choice, though a model must of course be included in the initial set of models if it is to be considered at all.
 
In theory, forecast accuracy is improved by taking relevant additional information into account, most often in the form of other data, such as in the case of joint forecasting of the fertility of a group of countries \citep{OP05}. In model averaging, given a single set of data, the additional information encompasses the features of multiple forecasting models. Model averaging can be expected to improve forecast accuracy because the strengths of different assumptions, model structures and degrees of model complexity are essentially combined through weighting that assigns greater weight to more accurate models. In other words, the model-averaged forecasts are more robust to model misspecification. 

This advantage might also be expected to increase with the forecast horizon. For any model, uncertainty increases with forecast horizon because the relevant experience in the increasingly distant fitting period becomes more and more attenuated. The effects of model misspecification may be cumulated over time, or increasingly accentuated. By using model averaging, the different effects of model misspecification are averaged rather than cumulated, potentially resulting in reduced uncertainty, especially when counterbalancing occurs, and curtailing the extent to which uncertainty increases with forecast horizon. Our results provide some evidence of a greater model-averaging advantage at longer horizons. The model-averaged point and interval forecasts gain ground as forecast horizon increases in terms of accuracy in relation to their six constituent forecasts (see Tables~\ref{tab:2} and~\ref{tab:3}).

Model averaging can also be expected to improve the reliability of forecast accuracy because as an average its range is limited. The model averaging approaches may not be the most accurate forecast among the set, but they are not likely to be the least accurate. In effect, model averaging imposes a lower bound on point and interval forecast accuracy. When the weights are derived empirically based on forecast accuracy from past data, model average accuracy can be expected to be relatively high and can be more accurate than any individual model. An example is seen in Table~\ref{fig:2}.

It is important to note that these advantages of model averaging hold for any set of models designed to predict period or cohort fertility behaviour at any level of detail. In other words, it is not a requirement that the forecasting models are purely extrapolative, as in this paper. Models based on other approaches such as expert judgement \citep[e.g.,][]{LSS+96} and explanation \citep[e.g.,][]{Ermisch92}, as well as Bayesian modelling \citep[e.g.,][]{SGM+14} can also be used in model averaging in order to take into account their different strengths in predicting fertility behaviour. Indeed, the traditional `medium'  fertility assumption could be included in the set of models. 

\subsection{Limitations}

A limitation of this study is that only six models from two families were considered. The set of models included in model averaging will influence the outcome. In this study, we included the RWD, although its appropriateness for longer-term fertility forecasting is limited because of changing trends \citep{MGC13} and potential inconsistencies arising from independent forecasting of age-specific rates \citep{Bell88}. The study confirms the RWD to be least accurate among the six models for both point and interval forecasts. Similarly, the RW is less appropriate for longer horizons because of changing trends. Despite these limitations of the selected models, gains in forecast accuracy were achieved through model averaging. With a larger set of models, greater gains may be achieved because a more diverse range of assumptions and model structures would be taken into account. 

A further limitation, arising from data availability, is the focus on period fertility rates, although it is period rates that are used in population projection. Cohort fertility rates have the advantage of describing the fertility behaviour of women over their life course, while parity-specific period rates would provide useful detail to improve model accuracy.  Unfortunately, lengthy annual time series of cohort data and parity-specific period data are not available for most of the selected countries, precluding their use in forecasting. 

\subsection{Future research}

The findings of this research suggest that model averaging is a potentially fruitful approach to improving the accuracy of fertility forecasting, using models that already exist. A useful technical extension in the case of the frequentist approach to model averaging, would be to develop a single set of weights based on a collective criterion incorporating both point and interval forecast accuracy. Additionally, a potential direction for further technical development is to examine other ways of combining models (bagging, boosting and stacking, for example) to achieve greater accuracy than any of the individual models \citep{BG69}. 

Based on our findings, we conclude that in the context of fertility forecasting the model-averaging approach warrants further examination using a more extensive range of models and considering both period and cohort data. The choice of potential models is extensive \citep{Booth06, BLM18} and could usefully include short-term models and those based on explanation and expert opinion. Examination of the  weights assigned to different models over time may help to increase understanding of fertility change. Further insights may be gained from the application of model averaging to forecasting total fertility \citep{Saboia77, OP05, ARG+11}, with comparisons of accuracy between direct and indirect forecasting of this aggregate measure. 
 
Finally, we note that optimal forecast accuracy is unlikely to be achieved by preselecting an approach or particular models, but rather is to be found in applying model averaging to a wide range of models encompassing differing strengths in predicting human behaviour. However, while model averaging can be expected to improve forecast accuracy beyond that of the individual models, it is also the case that the accuracy of the individual models broadly determines the range of accuracy that the model-averaged forecast might achieve. Thus, while the prospect of increased accuracy in fertility forecasting is enhanced by the potential of model averaging, the addition of model averaging to the forecasting toolset does not absolve demographers from striving to better understand and model fertility behaviour.

\newpage
\begin{appendices}
\section{Model averaging methods: technical details}\label{sec:MA_tech}

Model averaging involves the computation of weighted means for the point forecast and the interval forecast. Depending on the model-averaging method employed, the weights may be the same for point and interval forecasting, or they may differ. The derivation of weights is described below for the two more complex methods employed in this study, the Bayesian and model confidence set approaches. Full details of the frequentist approach and the use of equal weights appear in Sections~\ref{sec:3.1} and~\ref{sec:equal_weight}.

\subsection{A Bayesian viewpoint}

Among the models in the set, let one be considered the correct model, and let $\theta_1,\theta_2,\dots,\theta_L$ be the vector of parameters associated with each model. Let $\Delta$ be the quantity of interest, such as the combined forecast of ASFRs; its posterior distribution given the observed data, $D$, is
\begin{align*}
\text{Pr}(\Delta|D) &= \sum^L_{i=1}\text{Pr}(\Delta|M_{\ell},D)\text{Pr}(M_{\ell}|D) \\
&=\sum^L_{i=1}\text{Pr}(\Delta|M_{\ell},D)\frac{\text{Pr}(D|M_{\ell})\text{Pr}(M_{\ell})}{\sum^L_{l=1}\text{Pr}(D|M_l)\text{Pr}(M_l)},
\end{align*}
where $\text{Pr}(D|M_{\ell})=\int\text{Pr}(D|\theta_{\ell},M_{\ell})\text{Pr}(\theta_{\ell}|M_{\ell})d\theta_{\ell}$ and $\theta_{\ell}$ represents the parameters in $M_{\ell}$, $\text{Pr}(\theta_{\ell}|M_{\ell})$ is the prior density of $\theta_{\ell}$ under model $M_{\ell}$, $\text{Pr}(D|\theta_{\ell},M_{\ell})$ is the likelihood, and $\text{Pr}(M_{\ell})$ is the prior probability that $M_{\ell}$ is the true model.

Given diffuse (also known as non-informative) priors and equal model prior probabilities, the weights for Bayesian model averaging are approximately
\begin{align}
w_{\ell} = \tilde{w}_{\ell} = \frac{\exp(-\frac{1}{2}\text{BIC}_{\ell})}{\sum^L_{\ell=1}\exp(-\frac{1}{2}\text{BIC}_{\ell})}, \qquad \ell=1,\dots,L,\label{eq:1}
\end{align}
where $\text{BIC}_{\ell} = 2\mathcal{L}_{\ell}+\log(n)\eta_{\ell}$,  $\mathcal{L}_{\ell}$ is the negative of the log-likelihood, $\eta_{\ell}$ is the number of parameters in model $\ell$, $n$ represents the number of years in the fitting period, and BIC$_{\ell}$ is the Bayesian information criterion for model $\ell$ which can be used as a simple and accurate approximation of the log Bayes factor \citep[see also][Section 11.11]{KW95, KR95, Raftery95, BA02, BA04, Ntzoufras09}. In the case of least squares estimation with normally distributed errors, BIC can be expressed as
\begin{equation*}
\text{BIC}_{\ell} = n\log(\widehat{\sigma}^2) + \log(n)\eta_{\ell},
\end{equation*}
where $\widehat{\sigma}^2 = \sum^n_{i=1}(\widehat{\varepsilon}_i)^2/n$ and $(\widehat{\varepsilon}_1,\dots,\widehat{\varepsilon}_n)$ are the residuals from the fitted model. The $\text{BIC}_{\ell}$ values are not interpretable, as they contain arbitrary constants and are much affected by sample size. We thus adjust $\text{BIC}_{\ell}$ by subtracting the minimum $\{\text{BIC}_{\ell}; \ell=1,\dots,L\}$. The adjusted $\text{BIC}_{\ell}$ is used in~\eqref{eq:1}.

\subsection{Model confidence set (MCS)}

As the equal predictive ability (EPA) test statistic can be evaluated for any loss function, we adopt the tractable absolute error measure. The procedure begins with an initial set of models of dimension $L$ encompassing all the models considered, $M_0 = \{M_1, M_2, \dots, M_L\}$. For a given confidence level, a smaller set, the superior set of models $\widehat{M}_{1-\alpha}^*$ is determined where $m^*\leq m$. The best scenario is when the final set consists of a single model, i.e., $m=1$. Let $l_{\ell,t}$ denote forecast error for model $\ell$ at time $t$, and let $d_{\ell \tau,t}$ denote the loss differential between two models $\ell$ and $j$, that is
\begin{equation*}
d_{\ell \tau,t} = l_{\ell,t} - l_{\tau,t}, \qquad \ell, \tau=1,\dots,L, \quad t=1,\dots,n,
\end{equation*}
and calculate
\begin{equation*}
d_{\ell\cdot,t} = \frac{1}{L-1}\sum_{\tau\in \text{M}}d_{\ell\tau,t}, 
\end{equation*}
as the loss of model $\ell$ relative to any other model $\tau$ at time point $t$. The EPA hypothesis for a given set of $\text{M}$ models can be formulated in two ways:
\begin{align}
\text{H}_{\text{0,M}}: c_{\ell\tau}&=0, \qquad \text{for all}\quad \ell, \tau = 1,2,\dots,L\notag\\
\text{H}_{\text{A,M}}: c_{\ell\tau}&\neq 0, \qquad \text{for some}\quad \ell, \tau = 1,2,\dots,L.\label{eq:2}
\end{align}
or
\begin{align}
\text{H}_{\text{0,M}}: c_{\ell.}&=0, \qquad \text{for all}\quad \ell, \tau = 1,2,\dots,L\notag\\
\text{H}_{\text{A,M}}: c_{\ell.}&\neq 0, \qquad \text{for some}\quad \ell, \tau = 1,2,\dots,L.\label{eq:3}
\end{align}
where $c_{\ell\tau} = \text{E}(d_{\ell\tau})$ and $c_{\ell.} = \text{E}(d_{\ell.})$ are assumed to be finite and time independent. Based on $c_{\ell \tau}$ or $c_{\ell.}$, we construct two hypothesis tests as follows:
\begin{equation}
t_{\ell\tau} = \frac{\overline{d}_{\ell\tau}}{\sqrt{\widehat{\text{Var}}\left(\overline{d}_{\ell\tau}\right)}}, \qquad
t_{\ell.} = \frac{\overline{d}_{\ell.}}{\sqrt{\widehat{\text{Var}}\left(\overline{d}_{\ell.}\right)}}, \label{eq:4} 
\end{equation}
where $\overline{d}_{\ell.} = \frac{1}{L-1}\sum_{\tau\in \text{M}}\overline{d}_{\ell\tau}$ is the sample loss of $\ell^{\text{th}}$ model compared to the averaged loss across models in the set $M$, and $\overline{d}_{\ell\tau} =\frac{1}{n}\sum^n_{t=1}d_{\ell\tau,t}$ measures the relative sample loss between the $\ell^{\text{th}}$ and $\tau^{\text{th}}$ models. Note that $\widehat{\text{Var}}\left(\overline{d}_{\ell.}\right)$ and $\widehat{\text{Var}}\left(\overline{d}_{\ell\tau}\right)$ are the bootstrapped estimates of $\text{Var}\left(\overline{d}_{\ell.}\right)$ and $\text{Var}\left(\overline{d}_{\ell\tau}\right)$, respectively. From \cite{HLN11} and \cite{BC14}, we perform a block bootstrap procedure with 5,000 bootstrap samples, where the block length is given by the maximum number of significant parameters obtained by fitting an AR($p$) process on all the $d_{\ell\tau}$ term. For both hypotheses in~\eqref{eq:2} and~\eqref{eq:3}, there exist two test statistics:
\begin{equation}
T_{\text{R,M}} = \max_{\ell,\tau\in \text{M}}|t_{\ell\tau}|,\qquad T_{\max, \text{M}} = \max_{\ell\in \text{M}} t_{\ell}, \label{eq:5}
\end{equation} 
where $t_{\ell\tau}$ and $t_{\ell.}$ are defined in~\eqref{eq:4}.

The selection of the worse-performing model is determined by an elimination rule that is consistent with the test statistic,
\begin{equation*}
e_{\text{R,M}} = \argmax_{\ell\in M}\left\{\sup_{\tau\in M}\frac{\overline{d}_{\ell\tau}}{\sqrt{\widehat{\text{Var}}\left(\overline{d}_{\ell\tau}\right)}}\right\},\qquad e_{\max, \text{M}} = \argmax_{\ell\in M}\frac{\overline{d}_{\ell.}}{\widehat{\text{Var}}\left(\overline{d}_{\ell.}\right)}.
\end{equation*}

To summarise, the MCS procedure to obtain a superior set of models consists of the following steps:
\begin{enumerate}
\item[1)] Set $M=M_0$;
\item[2)] If the null hypothesis is accepted, then the final model $M^* = M$; otherwise use the elimination rules defined in~\eqref{eq:5} to determine the worst model;
\item[3)] Remove the worst model and go to Step 2). 
\end{enumerate}

In a particular case of the MCS approach, a lower confidence level results in the inclusion of all models without weights. With a higher confidence level, the MCS approach selects only the most accurate model.

\newpage
\section{Selected models: technical details}\label{sec:appendix_B}

We introduce the six models and describe the calculation of point and interval forecasts for ASFRs. In order to ensure non-negative predictions, prior to modelling, the ASFRs are first transformed using the Box-Cox transformation which is defined as
\[ m_t(x_i) = \left\{ \begin{array}{ll}
         \frac{1}{\kappa}\left([f_t(x_i)]^{{\kappa}-1}\right) & \mbox{if $0<\kappa \leq 1$};\\
        \ln[f_t(x_i)] & \mbox{if $\kappa = 0$}\end{array} \right. \quad i=1,2,\dots,p,\quad  t=1,2\dots,n, \]
where $f_t(x_i)$ represents the observed ASFR at age $x_i$ in year $t$, $m_t(x_i)$ represents the transformed ASFR, and $\kappa$ is  the transformation parameter.  Following \cite{HB08} and \cite{Shang15b}, we use $\kappa = 0.4$ as it gave relatively small holdout sample forecast errors on the untransformed scale.

\subsection{Hyndman-Ullah (HU) model and two variants}

The HU model \citep{HU07} is described below, along with the defining features of its two variants, the robust HU model (HUrob) and the weighted HU model (HUw). 
\begin{asparaenum}
\item[1)] The transformed ASFRs are first smoothed using a concave regression spline \citep[see][for details]{HU07, SCA16}. We assume an underlying continuous and smooth function $\{s_t(x); x\in [x_1,x_p]\}$ that is observed with error at discrete ages,
\begin{equation*}
m_t(x_i) = s_t(x_i) + \sigma_t(x_i)\varepsilon_{t,i}, \qquad \text{for}\quad i=1,2,\dots,p,\quad  t=1,2\dots,n,
\end{equation*}
where $\sigma_t(x_i)$ allows the amount of noise to vary with $x_i$ in year $t$, and $\varepsilon_{t,i}$ is an independent and identically distributed (iid) standard normal random variable. As an example, Figure~\ref{fig:4} displays the original and smoothed Australian ASFRs.
\begin{figure}[!htbp]
  \centering
  \subfloat[Observed ASFR time series]
  {\includegraphics[width=8.8cm]{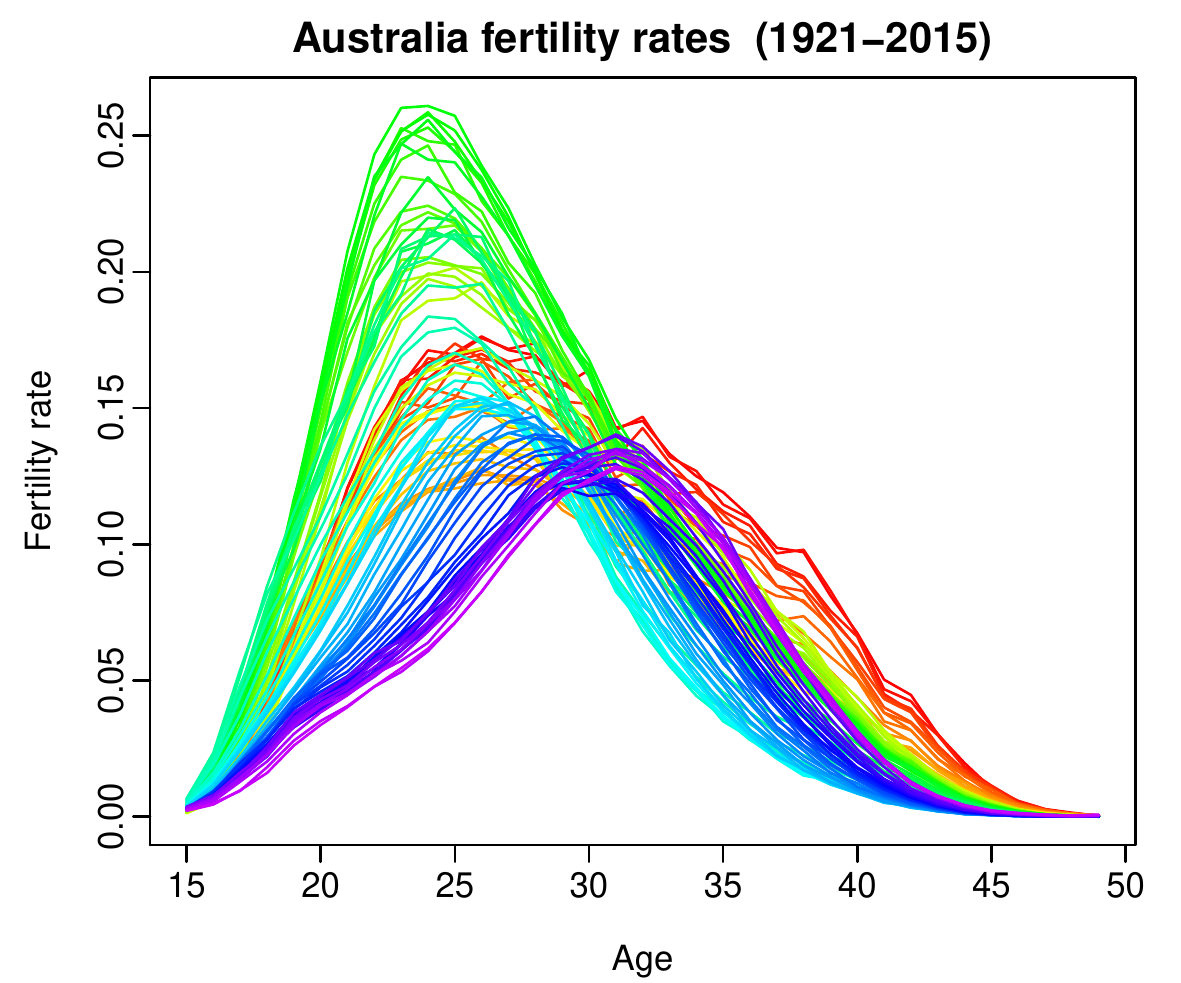}}
  \qquad
  \subfloat[Smoothed ASFR time series]
  {\includegraphics[width=8.8cm]{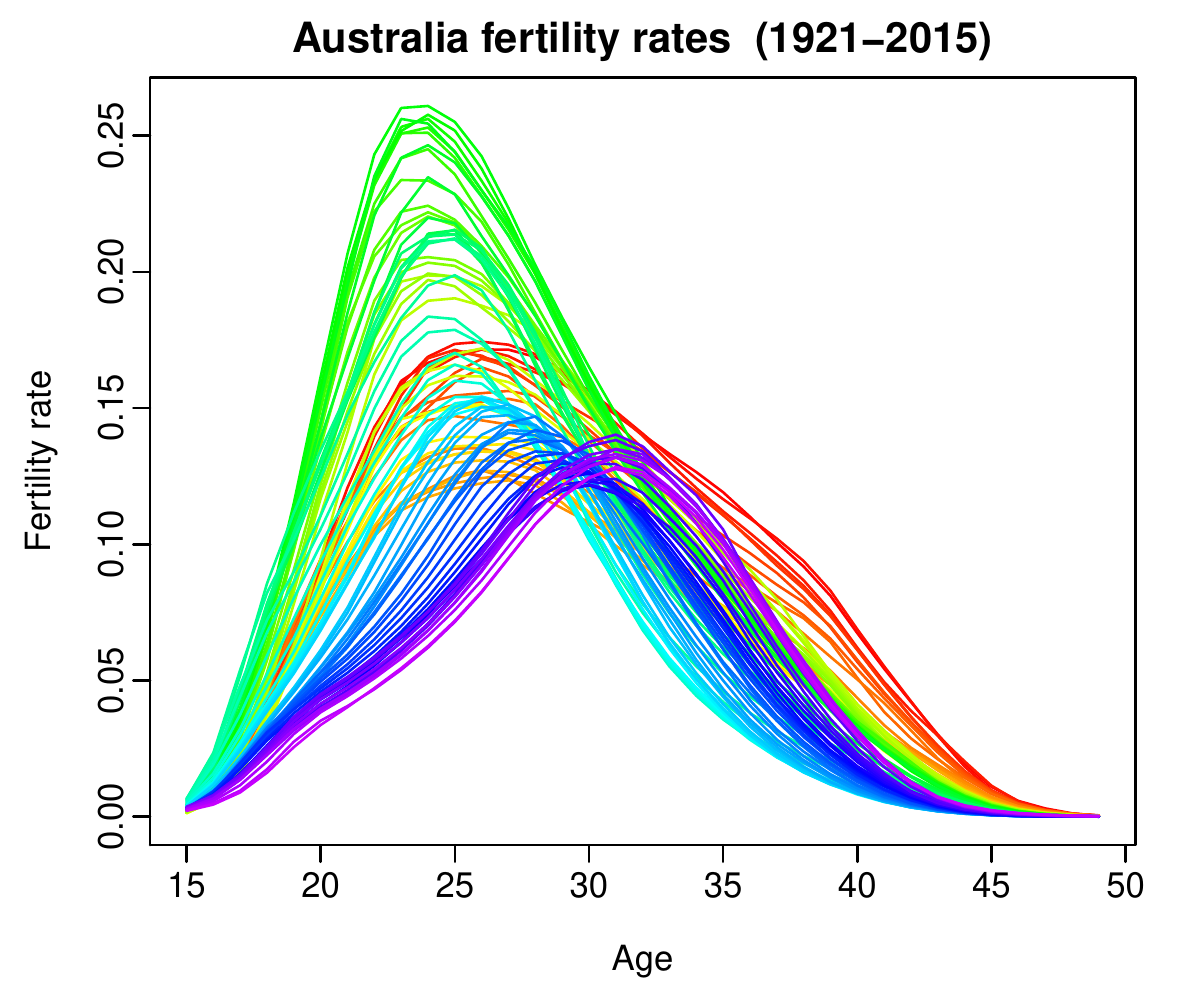}}
\caption{Observed and smoothed ASFRs for Australia, 1921 to 2015. The data are represented chronologically by the colours of the rainbow from red to violet (most recent)}\label{fig:4}
\end{figure}
\item[2)] Given the set of continuous and smooth curves $\{s_1(x),s_2(x),\dots,s_n(x)\}$, the mean function $a(x)$ is estimated by
\begin{equation*}
\widehat{a}(x)=\sum^n_{t=1}\varpi_ts_t(x),
\end{equation*}
where $\varpi_t=1/n$ represents equal weighting in the HU and HUrob models. In the HUw model, $\{\varpi_t=\lambda(1-\lambda)^{n-t},t=1,2,\dots,n\}$ represents a set of geometrically-decaying weights \citep[see][on the selection of optimal $\lambda$]{HS09}. The distinguishing feature of the HUw model is that forecasts are based more heavily on recent data.
\item[3)] Using functional principal components analysis, the set of continuous and smooth curves $\left\{s_t(x);t=1,2,\dots,n\right\}$ is decomposed into orthogonal functional principal components and their uncorrelated scores:
\begin{equation*}
s_t(x) = \widehat{a}(x) + \sum^J_{j=1}\widehat{b}_j(x)\widehat{k}_{t,j}+\widehat{e}_t(x),
\end{equation*}
where $\left\{\widehat{b}_1(x),\widehat{b}_2(x),\dots,\widehat{b}_J(x)\right\}$ represents a set of weighted (as above) functional principal components; $\left\{\widehat{k}_{t,1},\widehat{k}_{t,2},\dots,\widehat{k}_{t,J}\right\}$ is a set of uncorrelated scores; $\widehat{e}_t(x)$ is the estimated error function with mean zero; and $J<n$ is the number of retained functional principal components. Following \cite{HB08}, we use $J=6$ as this has been shown to be sufficiently large to produce iid residuals with a mean of zero and finite variance.

The HUrob model uses $\widehat{e}_t(x)$ to provide information about possible outlying curves. If a curve has a more considerable value of integrated square error, it indicates that this curve may be considered as an outlier that generates from a different data generating process than the rest of the observations. An example of this is fertility rates during the baby boom when sharp temporal increases occurred. Computationally, we use the hybrid algorithm of \cite{HU07} with 95\% efficiency (where the most outlying 5\% of the data are removed).
\item[4)] By conditioning on the observed data $\bm{\mathcal{I}}=\left\{m_1(x_i),\dots,m_n(x_i)\right\}$ and the set of estimated functional principal components $\bm{B}=\left\{\widehat{b}_1(x),\dots,\widehat{b}_J(x)\right\}$, the $h$-year-ahead point forecast of $m_{n+h}(x)$ can be expressed as
\begin{equation*}
\widehat{m}_{n+h|n}(x)=\text{E}[m_{n+h}(x)|\bm{\mathcal{I}},\bm{B}]=\widehat{a}(x)+\sum^J_{j=1}\widehat{b}_j(x)\widehat{k}_{n+h|n,j},
\end{equation*}
where $\widehat{k}_{n+h|n,j}$ denotes the $h$-year-ahead forecast of $k_{n+h,j}$ using a univariate time-series model, such as an optimal autoregressive integrated moving average (ARIMA) model (see Section~\ref{sec:optimal}). Note that multivariate time-series forecasting models, such as vector autoregressive and vector autoregressive moving average models, can also be used to forecast principal component scores \citep[see, e.g.,][]{ANH15}. 

Because of the orthogonality of the functional principal components, the overall forecast variance can be approximated by the sum of four variances:
\begin{equation}
\text{Var}[m_{n+h}(x)|\bm{\mathcal{I}},\bm{B}]\approx \widehat{\sigma}_a^2(x)+\sum^J_{j=1}[\widehat{b}_j(x)]^2u_{n+h|n,j}+v(x)+\sigma_{n+h}^2(x), \label{eq:6}
\end{equation}
where $\widehat{\sigma}_a^2(x)$ is the variance of the mean function, estimated from the difference between the sample mean function and the smoothed mean function; $\left[\widehat{b}_j(x)\right]^2u_{n+h|n,j}$ is the variance of $j$th estimated functional principal component decomposition where $u_{n+h|n,j}=\text{Var}\left(k_{n+h,j}|\widehat{k}_{1,j},\dots,\widehat{k}_{n,j}\right)$ can be obtained from the univariate time-series models used for forecasting scores; $v(x)$ is the model error variance estimated by averaging $\{\widehat{e}^2_1(x),\dots,\widehat{e}_n^2(x)\}$ for each $x$, and $\sigma_{n+h}^2(x)$ is the smoothing error variance estimated by averaging $\{\widehat{\sigma}_1^2(x),\dots,\widehat{\sigma}_n^2(x)\}$ for each $x$. The prediction interval is constructed via~\eqref{eq:6} on the basis of normality.
\end{asparaenum}

\subsection{Univariate time-series models}

The three selected univariate time-series models \citep{BJR08} are applied to transformed ASFRs at each age. 

\paragraph{Random-walk models}

For each age $x_i$, the random walk (RW) and random walk with drift (RWD) models are
\begin{equation*}
m_{t+1}(x_i)=c+m_{t}(x_i)+e_{t+1}(x_i),
\end{equation*}
where $c$ represents the drift term capturing a possible trend in the data (in the RW model $c=0$), and $e_{t+1}(x_i)$ represents the iid normal error with a mean of zero.

The $h$-year-ahead point forecast and total variance used for constructing the prediction interval are given by
\begin{align*}
\widehat{m}_{n+h|n}(x_i)&=\text{E}[m_{n+h}(x_i)|m_{1}(x_i),\dots,m_{n}(x_i)]=ch+m_{n}(x_i),\\
\text{Var}[\widehat{m}_{n+h|n}(x_i)]&= \text{Var}[m_{n+h}(x_i)|m_{1}(x_i),\dots,m_{n}(x_i)]=\text{Var}[m_{n}(x_i)]+\text{Var}[e_{n+h}(x_i)],
\end{align*}
where $h$ represents the forecast horizon.

\paragraph{Optimal ARIMA model}\label{sec:optimal}

An ARIMA$(p,d,q)$ model has autoregressive components of order $p$ and moving average components of order $q$, with $d$ being the degree of difference needed to achieve stationarity \citep{BJR08}. The model can be expressed as
\begin{equation*}
\Delta^dm_{t}(x_i) = c+\sum^p_{\upsilon=1}\beta_{\upsilon}\Delta^dm_{t-\upsilon}(x_i)+\omega_{t}(x_i)+\sum^q_{j=1}\psi_j\omega_{t-j}(x_i),\qquad t=\max(p,q)+1,\dots,n,
\end{equation*}
where $\Delta^dm_{t}(x_i)$ represents the stationary time series after applying the difference operator of order $d$, $c$ is the drift term; $\{\beta_1,\dots,\beta_p\}$ represents the coefficients of the autoregressive components; $\{\psi_1,\dots,\psi_q\}$ represents the coefficients of the moving average components; and $\omega_{t}(x_i)$ is a sequence of iid random variables with mean zero and variance $\sigma_{\omega}^2$.

We use the \verb auto.arima  algorithm \citep{HK08} in the \verb forecast  package \citep{Hyndman20} to select the optimal orders based on the corrected Akaike information criterion, and then estimate the parameters by maximum likelihood. The one-year-ahead point and interval forecasts are given by
\begin{align}
\Delta^d\widehat{m}_{n+1|n}(x_i) = \text{E}[m_{n+1}(x_i)|m_{1}(x_i),\dots,m_{n}(x_i)]&=c+\sum^p_{\upsilon=1}\widehat{\beta}_{\upsilon}\Delta^dm_{n+1-\upsilon}(x_i),\label{eq:7} \\
\text{Var}[m_{n+1}(x_i)|m_{1}(x_i),\dots,m_{n}(x_i)]&=\widehat{\sigma}_{\omega}^2\left(1+\widehat{\psi}_1^2+\dots+\widehat{\psi}_q^2\right).\label{eq:8}
\end{align}
For the $h$-year-ahead forecasts,~\eqref{eq:7} and~\eqref{eq:8} are applied iteratively.

\newpage
\section{Forecast accuracy evaluation}\label{sec:5}

\subsection{Evaluation of point forecast accuracy}\label{sec:5.1}

Following \cite{BHT+06} and \cite{SBH11}, we use MAFE to measure point forecast accuracy. MAFE is the average of absolute error, $|\text{actual}-\text{forecast}|$, across countries (in this study, $g=1,\dots,17$), years in the forecasting period and ages; it measures forecast precision regardless of sign. 
\begin{equation*}
\text{MAFE}_h = \frac{1}{17\times (21-h)\times 35}\sum^{17}_{g=1}\sum^{n+20-h}_{r=n}\sum^{35}_{i=1}\left|m_{r+h}(x_i)-\widehat{m}_{r+h|r}(x_i)\right|,
\end{equation*}
where $h$ denotes forecast horizon, $m_{r+h|r}(x_i)$ represents the actual ASFR at age $x_i$ in the forecasting period, and $\widehat{m}_{r+h}(x_i)$ represents the forecast. Note that $x_1$ corresponds to age 15, and $x_{35}$ corresponds to age 49.

\subsection{Evaluation of interval forecast accuracy}\label{sec:5.2}

To assess interval forecast accuracy, we use the interval score of \cite{GR07} \citep[see also][]{GK14}. For each year in the forecasting period, one-year-ahead to 20-year-ahead prediction intervals were calculated at the $80\%$ nominal coverage probability, with lower and upper bounds that are predictive quantiles at $10\%$ and $90\%$. As defined by \cite{GR07}, a scoring rule for the interval forecast at age $x_i$ is expressed as
\begin{align*}
S_{\alpha}& \left[\widehat{m}_{r+h|r}(x_l),\widehat{m}_{r+h|r}(x_u);m_{r+h}(x_i)\right] = \left[\widehat{m}_{r+h|r}(x_u)-\widehat{m}_{r+h|r}(x_l)\right]+\frac{2}{\alpha}\left[\widehat{m}_{r+h|r}(x_l)-m_{r+h}(x_i)\right]\\
& \mathds{1}\left\{m_{r+h}(x_i)<\widehat{m}_{r+h|r}(x_l)\right\}+ \frac{2}{\alpha}\left[m_{r+h}(x_i)-\widehat{m}_{r+h|r}(x_u)\right]\mathds{1}\left\{m_{r+h}(x_i)>\widehat{m}_{r+h|r}(x_u)\right\},
\end{align*}
where $\alpha$ (customarily 0.2) denotes the level of significance, $\widehat{m}_{r+h|r}(x_l)$ and $\widehat{m}_{r+h|r}(x_u)$ represent the lower and upper prediction intervals at age $x_i$ in the relevant forecasting period and country. The interval score rewards a narrow prediction interval, if and only if the true observation lies within the prediction interval. The optimal score is achieved when $m_{r+h}(x_i)$ lies between $\widehat{m}_{r+h|r}(x_l)$ and $\widehat{m}_{r+h|r}(x_u)$, and the distance between $\widehat{m}_{r+h|r}(x_l)$ and $\widehat{m}_{r+h|r}(x_u)$ is minimal. 

The mean interval score is averaged over age. For multiple  countries (here, 1 to 17) and years in the forecasting period (here, 1 to 20), the mean interval score is further averaged:
\begin{small}
\begin{equation*}
\overline{S}_{\alpha} \left[\widehat{m}_{r+h|r}(x_l),\widehat{m}_{r+h|r}(x_u);m_{r+h}(x_i)\right]  = \frac{1}{17\times (21-h)\times 35}\sum^{17}_{g=1}\sum^{n+20-h}_{r=n}\sum^{35}_{i=1}S_{\alpha} \left[\widehat{m}_{r+h|r}(x_l),\widehat{m}_{r+h|r}(x_u);m_{r+h}(x_i)\right].
\end{equation*}
\end{small}
\end{appendices}

\newpage
\bibliographystyle{agsm}
\bibliography{fertility}

\end{document}